\def\XXint#1#2#3{{\setbox0=\hbox{$#1{#2#3}{\int}$}
     \vcenter{\hbox{$#2#3$}}\kern-.5\wd0}}

\documentclass[aps,prl,twocolumn,showpacs,floatfix]{revtex4-1}

\pdfoutput=1

\usepackage{times}
\usepackage{amsfonts}
\usepackage{amssymb}
\usepackage{amsmath}
\usepackage{graphicx}
\usepackage{bm}
\usepackage{verbatim}

\usepackage{hyperref}

\usepackage{color}
\usepackage{stackrel}
\usepackage{accents}

\usepackage{latexsym}

\usepackage[cspex,bbgreekl]{mathbbol}

\usepackage{centernot}

\newcommand{\bsub}{\begin{subequations}}
\newcommand{\esub}{\end{subequations}}

\newcommand \bea {\begin{eqnarray} }
\newcommand \eea {\end{eqnarray}}
 
\newcommand{\beg}{\begin{equation}}
\newcommand{\en}{\end{equation}}

\newcommand \bel  {\begin{align}}
\newcommand \enl  {\end{align}}

\newcommand{\eps}{\varepsilon}
\newcommand{\lam}{\lambda}
\newcommand{\ph}{\varphi}

\newcommand{\re}[1]{(\ref{#1})}

\newcommand{\eref}[1]{Eq.~(\ref{#1})}

\newcommand{\Tr}{\mathrm{Tr}\,}

\newcommand{\pmat}{\begin{pmatrix}}
\newcommand{\epmat}{\end{pmatrix}}

\def\8{\infty}

\def\undertext#1{\vtop{\hbox{#1}\kern 1pt \hrule}}

\def\be{\begin{equation}}
\def\ee{\end{equation}}
\def\bea{\begin{eqnarray} & &}
\def\eea{\end{eqnarray}}


\begin{document}

\title{Generalized microcanonical and Gibbs ensembles in classical and quantum integrable dynamics}

\author{Emil A. Yuzbashyan }
\affiliation{Center for Materials Theory, Rutgers University, Piscataway, NJ 08854, USA\\
}

\begin{abstract}

We prove two statements about the long time dynamics of integrable Hamiltonian systems. In classical mechanics, we   prove the microcanonical version of the Generalized Gibbs Ensemble (GGE)  by mapping it to a known theorem    and then  extend it to the limit of infinite number of degrees of freedom.
In quantum mechanics, we prove GGE for maximal Hamiltonians -- a class of models stemming from a  rigorous notion of quantum integrability understood  as the existence of    conserved charges with prescribed dependence on a  system parameter, e.g. Hubbard $U$, anisotropy in the XXZ model etc.  In analogy with classical integrability, the defining property of  these models is that they have   the maximum   number of independent integrals. We contrast their dynamics   induced by quenching the parameter to that of random matrix Hamiltonians.

\end{abstract}

\maketitle

The past decade has witnessed an unprecedented experimental access to global coherent dynamics of many-body interacting systems. As a result, a new area  that could be called  ``far from equilibrium many-body Hamiltonian dynamics", ``coherent many-body dynamics" or  ``quantum quenches" has emerged. A major part of research in this area has focused on testing the GGE\cite{rigol, tolya}  in various integrable models. GGE refers to a density matrix or, in the case of classical mechanics, a phase space distribution  function
\beg
\rho=Z^{-1} e^{-\sum_k \beta_k H_k},
\label{gge}
\en
where $H_k$ are a (complete in some sense) set of integrals of motion for system Hamiltonian $H$ and $Z$ is a normalization constant. Suppose the system evolves  with $H$ starting from a non-stationary state. The statement of GGE is that the infinite time average of an observable $O$ coincides with its ensemble average with the density matrix $\rho$\cite{note0}.  

Most authors test  GGE in quantum models without clarifying their notion of quantum integrability. The latter however is a tricky concept with no  generally accepted definition,  making the quantum GGE conjecture essentially  unfalsifiable. The notion of classical integrability on the other hand is unambiguous\cite{arnold}. For this and other reasons, it makes sense to  first understand the status of GGE  in classical mechanics. We will see   that the microcanonical version of GGE -- Generalized Microcanonical Ensemble --   is  exact for a general classical integrable Hamiltonian. In a parallel line of inquiry, we  will prove  GGE for a  class of models that emerge from  a recently proposed complete notion of quantum integrability.

Generalized Microcanonical Ensemble (GME) in classical mechanics is the following phase space distribution:
\beg
\rho (\bm p, \bm q)= V^{-1}\prod_{k=1}^n \delta\left(H_k(\bm p, \bm q)-h_{k}\right),
\label{gme}
\en
where  $\bm q=(q_1,\dots,q_n)$ and $\bm p=(p_1,\dots,p_n)$ are the generalized coordinates and momenta.
Suppose the system evolves with an integrable Hamiltonian $H(\bm p,\bm q)$ starting  from a point $(\bm p_0, \bm q_0)$. Let $h_k=H_k(\bm p_0, \bm q_0) $ be the values of its integrals of motion for this initial condition. The statement of GME is that the time average of any dynamical variable  $O(\bm p,\bm q)$ is equal to its phase space average with distribution~\re{gme},
\beg
\lim_{T\to\infty} \frac{1}{T}\int_0^T\!\! O\left(t\right) dt=\int \!\! O(\bm p, \bm q)\rho(\bm p,\bm q)d\bm pd\bm q,
\label{thm}
\en
where $O\left(t\right)= O(\bm p(t), \bm q(t))$. \eref{thm} also holds for integrable classical spin Hamiltonians $H(\{\vec s_k\})$, in which case $p_k=\cos\theta_k$ and $q_k=\phi_k$, where $\theta_k$ and $\phi_k$ are the polar and azimuthal angles defining the spin direction. 
\eref{thm} is valid for any number of degrees of freedom $n$, so one can take the limit $n\to\infty$ on both sides. Moreover, we will argue that the limits $n\to\infty$ and $T\to\infty$  commute (a tremendous simplification) as long as  the frequency spectrum of $O(t)$ is free from a certain anomaly near the zero frequency.

As a first step towards a similarly unambiguous  statement in quantum mechanics, we also  analyze GGE in the framework of a rigorous formulation of quantum integrability\cite{owusu}.
Simplest  models  that arise in this approach are  type-1 or maximal Hamiltonians -- general $N$ linearly independent   commuting $N\times N$ Hermitian matrices of the form $H(x) = T+xV$,
where $x$ is a real parameter. Type-1 matrices represent blocks of various exactly solvable many-body models (such as 1D Hubbard and Gaudin magnets)  for certain   sets of quantum numbers (total spin projection  etc.)\cite{owusu,owusu1,yuzbashyan,shastry} and also describe e.g. a short range impurity in a metallic grain\cite{aleiner}. We prove  GGE is exact   for any $N$ and explicitly determine $\beta_k$ in \eref{gge}. 
 The GGE density matrix for quenches of the parameter $x$  turns out to be   non-thermal for type-1 Hamiltonians. In contrast, if we choose  $T$ and $V$  randomly, the post-quench asymptotic state is thermal in $N\to\infty$ limit. This emphasizes the importance of  a well-defined notion of integrability as naively one could claim $N$ integrals of motion (e.g. projectors onto the eigenstates) in the random matrix example too. We also relate the non-thermal behavior  to  localization. 
 
A characteristic feature of   type-1 and classical integrable systems  is that in both cases   the   number of independent integrals is the maximum allowed by the definition. Their dynamics are  constrained by the integrals apart from linear in time phases (angles) that  cancel out upon time-averaging or dephase in the thermodynamic limit. As the result, the integrals of motion fully determine infinite time averages. The situation when the number of conservation laws is appreciably less than the maximum is unclear and we will not consider it here.
A common belief is that in quantum exactly solvable systems this number  scales   as the logarithm of the size of the total Hilbert space, which
  agrees with most existing constructions of conserved charges (integrals). Further analysis however   reveals additional integrals\cite{owusu1,prosen}, which can be crucial in identifying the  proper    ensemble\cite{caux}. It therefore seems likely that some version of GGE or GME does generally hold in quantum mechanics if integrability is properly defined and all independent integrals are taken into account.

GGE apparently holds for relatively simple quantum models, such as 1D hard-core bosons\cite{rigol} and Luttinger liquids\cite{Cazalilla}, but fails e.g. in the XXZ and attractive Lieb-Liniger models for a seemingly reasonable choice of   integrals $H_k$\cite{wouters,pozsgay,goldstein}. More generally,  \cite{wouters,goldstein} argue that GGE  fails for models with bound states and \cite{gurarie}, that it reproduces global observables only in models mappable to noninteracting uncorrelated fermions. Given that \eref{thm} is a  rigorous theorem in classical mechanics, a natural question  is: what are  the  reasons for such failure? First, the equivalence between GME and GGE can break down already on the classical level. The standard argument to go from the microcanonical to the canonical ensemble requires that energy be an extensive property and interactions, roughly speaking, short ranged\cite{ruelle}. It is straightforward to extend this argument to the generalized ensembles, but then each $H_k$ must have these properties, which is not necessarily the case. For example, \textit{classical} Gaudin magnets  are well-defined integrable models with long ranged $H_k$.

Another set of problems arise from relying on an incomplete definition of quantum integrability. Main issues here are how we understand the independence  (nontriviality)   and   completeness of a set of integrals. Classical integrability, for example, requires $n$ functionally independent integrals.  If we allow all integrals in quantum GGE, one can simply choose the projectors onto the eigenstates of the Hamiltonian. With this choice of $H_k$ \eref{gge}
is equivalent to the diagonal ensemble for any Hamiltonian and the statement of GGE  becomes  tautological.  One might object that projectors are nonlocal (not short ranged). First, this is not the case in models with localized eigenstates. More importantly, while locality might be necessary for going from the microcanonical to the canonical ensemble, it hardly is  a legitimate requirement in the \textit{definition} of quantum integrability. Indeed, there is no such condition in the classical case and, moreover, there are quantum Hamiltonians, e.g. quantum Gaudin and BCS  models\cite{gaudin, cambiaggio} that are nonlocal, but otherwise bear all hallmarks of integrability. On the other hand, we do not expect GGE to hold for an incomplete set of integrals, at least the theorem \re{thm}  does not.

Both \eref{gge} and quantum infinite time average are invariant with respect to the choice of  any nondegenerate Hermitian operator within a given \textit{integrable family} as the system Hamiltonian. At this point we loosely define an integrable family as the set of operators that share the same  integrals of motion $H_k$. For example, in the usual construction of the conserved charges for Lieb-Liniger, 1D Hubbard, and XXZ models one of the $H_k$ is the Hamiltonian, while the rest  serve as its integrals. We can alternatively designate any other $H_k$ or their linear combination   as the  Hamiltonian without modifying \eref{gge} and the time average. Interestingly, there is  a combination, 
\beg
 H_\beta=\beta^{-1}\sum_k\beta_k H_k, 
\label{hb}
\en
for which GGE coincides with the Gibbs ensemble.  \eref{gme} is similarly independent of the choice of a nondegenerate  Hamiltonian (see below) within the classical integrable family.

To see the above invariance, note that the time average is given by the diagonal ensemble\cite{tolya}, 
\beg
\left\langle O(t)\right\rangle_t\equiv\lim_{T\to\infty} \frac{1}{T}\int_0^T\!\! \langle O\left(t\right)\rangle dt=\sum_m |c_m|^2 O_{mm},
\label{diag}
\en
where $c_m$ are the coefficients in the decomposition of the initial state $|\mathrm{in}\rangle$ into the eigenstates, which are  shared by all $H_k$. Similarly, conditions $\langle\mathrm{in}|H_k |\mathrm{in}\rangle=\Tr{\rho H_k}$ that determine $\beta_k$ are the same.
  Let us also note that \eref{diag} is useful for a macroscopic system  when  the thermodynamic and $T\to\infty$ limits commute. Arguments   we make below about the order of $n\to\infty$ and $T\to\infty$ limits in \eref{thm} apply here as well.

In what follows we first derive GGE for the maximal (type-1) Hamiltonians and then prove \eref{thm}. The approach of \cite{owusu,owusu1,yuzbashyan,shastry} to quantum integrability addresses many-body Hamiltonians at the level of blocks (sectors) stripped of all space-time and internal space symmetries. Basic objects are $N\times N$  Hermitian matrices
of the form $H(x)=T+xV$, where $x$ is a real parameter (Hubbard $U$, anisotropy in the XXZ model, magnetic field in Gaudin magnets etc.)   We say that $H(x)$ is integrable if it has at least one commuting partner  $\widetilde{H}(x)=\widetilde{T}+x\widetilde{V}$ other than a linear combination of $H(x)$ and the identity. An \textit{integrable family}   is a set
of     commuting, linearly independent  Hermitian matrices $H_i(x)=T_i+xV_i$, the most general Hamiltonian in the family being $H(x)=\sum_i d_i H_i(x)$. Thus, we define integrability simply as the existence of integrals linear in the parameter\cite{note2}.  

There is a natural classification of integrable families     by the number of independent commuting matrices they contain. We say that $H(x)$ is type-$M$ when  this number is $N-M+1$.  The maximum number of linearly independent, linear in $x$ integrals  $H(x)$ can have is $N$. Then it is a type-1 or, equivalently, a maximal Hamiltonian. 
It turns out that \textit{any} type-1  Hamiltonian can be parametrized by $3N$ numbers $d_i, \eps_i, \gamma_i$ as $H(x)=N^{-1}\sum_i d_i H_i(x)$, where
\beg
H_i(x)=  Np_i+xN\sum_{j\ne i} \frac{\gamma_i\gamma_j^* p_{ij}-|\gamma_j|^2 p_i-|\gamma_i|^2 p_j }{\eps_i-\eps_j},
\label{type1}
\en
 $p_{ij}=|i\rangle\langle j|+|j\rangle\langle i|$,  $p_j=|j\rangle\langle j|$,  and $|j\rangle$ are the shared eigenstates
of $T_i$.  Conversely, given arbitrary $2N$ real $d_i, \eps_i$ and $N$ complex $\gamma_i$, \eref{type1} yields a type-1 Hamiltonian $H(x)$.

 Normalized eigenstates $|\lam_m\rangle$ of $H_i(x)$ read
\beg
\langle i|\lam_m\rangle=\frac{\gamma_i}{{\cal N}_m(\lam_m-\eps_i)},\quad {\cal N}^2_m=\sum\limits_{k} \frac{|\gamma_k|^2}{(\lam_m-\eps_k)^2},
\label{states}
\en
where $\lam_m$ is any of the $N$ real roots of the  equation
\beg
\sum_{k} \frac{|\gamma_k|^2}{\lam_m-\eps_k}=\frac{1}{x}.
\label{roots}
\en
Eigenvalues of  general  type-1 $H(x)=N^{-1}\!\sum_k d_k H_k(x)$ are
\beg
E_m=x\sum_{k} \frac{d_k|\gamma_k|^2}{\lam_m-\eps_k}.
\label{Em}
\en
For example, $d_k=\eps_k$ yields an interesting Hamiltonian that describes a short range impurity in a metallic grain\cite{aleiner}
\beg
H_{\mathrm{im}}(x)=x|\gamma\rangle\langle\gamma|+\sum_i\eps_i |i\rangle\langle i|,
\en
with eigenvalues $E_m=\lam_m$. Here $ |\gamma\rangle=\sum_i\gamma_i|i\rangle$.

The number of integrals for type-1 is maximal and equals the dimension of the Hilbert space $N$. We can show that GGE is exact for any $N$ by matching the diagonal ensemble, i.e.
\beg
\langle \lam_m| e^{-\sum_k \beta_k H_k(x)}|\lam_m\rangle=Z|c_m|^2.
\en
With the help of \eref{Em} this becomes
\beg
x\sum_k\frac{\beta_k|\gamma_k|^2}{\lam_m-\eps_k}=\ln|c_m|^2+\ln Z,
\label{b}
\en
with  an explicit solution  
\beg
\beta_k=  \ln Z  +\frac{1}{x}\sum_m\frac{\ln|c_m|^2}{{\cal N}_m^2 (\lam_m-\eps_k)}.
\en
 An overall shift $\beta_k\to\beta_k +c$ adds a constant $c$ to the RHS of \eref{b} in view of \eref{roots}. The solution for $\beta_k$ is unique apart from the shift, i.e. an arbitrary choice of $Z$.
 
 To see how type-1  differs from a general $H(x)$ of the same form $T+xV$  (which has no nontrivial commuting partners in our definition), let us compare parameter quenches from $x_i=0$ to $x_f$ in type-1 Hamiltonians to those for random $T$ and $V$. Let the initial state be an eigenstate $ |i\rangle$ of $T$. 
 
 The diagonal ensemble for type-1 according to \eref{states} is
 \beg
 |c_m|^2=\frac{|\gamma_i|^2}{{\cal N}^2_m(\lam_m-\eps_i)^2}.
 \en
 A natural choice for $\eps_k$ are eigenvalues of a random matrix from the Gaussian orthogonal ensemble (GOE) with mean level spacing $\delta\propto N^{-1}$ and we also set $|\gamma_i|^2=N^{-1}$. \eref{roots} implies $\eps_{k-1}<\lambda_k<\eps_{k}$. It follows that in $N\to\infty$ limit $|c_m|^2\propto N^0$ for $\lam_m$ in an infinitesimal vicinity  of $\eps_i$ and $\propto N^{-2}$ otherwise. Note that this indicates localization of the eigenstate $|\lambda_m\rangle$ of $H(x)$ in the space of eigenstates of $T$\cite{ossipov}. For simplicity, we assume that the matrix element
 $\langle\lam_m|O|\lam_m\rangle\equiv f_O(\lam_m)$ is a smooth function of $\lam_m$ ($\lam_m$ becomes a continuous real variable when $N\to\infty$). \eref{diag} then yields $\langle O(t)\rangle_t = f_O(\eps_i)$.
 Consider next the GGE with $H_\beta=H_i(x)$, i.e. $\beta_k=\beta\delta_{ik}$. Its eigenvalues are $E_m^i=(\lam_m-\eps_i)^{-1}$. The ground state corresponds to $m=i$ and is separated by a gap $\propto N$ from  the excited states. The normalized density matrix is  simply $\rho_{mm}=\delta_{mi}$ and $\Tr(\rho O)=f_O(\eps_i)=\langle O(t)\rangle_t $. This GGE is non-thermal for any type-1 Hamiltonian   other than $H_i$, e.g. for $H_\mathrm{im}(x)$.
 
Now suppose $V$ is  also random and uncorrelated with $T$.  Eigenstates of $H(x)=T+xV$ and $T$   decorrelate at $x=O(1)$\cite{wilkinson}, so that $c_m$ are components of a random vector.  Averaged over  narrow energy windows $\langle|c_m|^2\rangle_E=N^{-1}$ at large $N$, which corresponds to the infinite temperature Gibbs distribution. Note that already in type-1 $T$ (or $V$) is arbitrary. A natural choice of $T$ is a GOE random matrix. However, any choice of $T$ severely constrains $V$ to ensure the existence of commuting partners.  It is precisely this correlation between $V$ and $T$ that also makes the density matrix non-thermal and eigenstates localized. Therefore, even though for a random $H(x)$ one can take $H_k$ in \eref{gge} to be the projectors onto its eigenstates, it is of no consequence because it does not introduce correlations between $V$ and $T$ and so constructed GGE is  just the Gibbs distribution,  i.e. $H_\beta=H(x)$. From this point of view, the statement of GGE is not  that \eref{gge} reproduces $\langle O(t)\rangle_t$, but that $H_\beta$ is distinct from $H$. Note that 
the exponential form of $\rho$ in \eref{gge} is unimportant  at finite $N$. It could as well be a different function of $H_\beta$ and we  still would be able to match the diagonal ensemble. It   however plays an important role in showing that $H_\beta\ne H$ in $N\to\infty$ limit, even though type-1 Hamiltonians are nonadditive. 

Now we switch gears to classical mechanics to prove \eref{thm}. There are two necessary conditions: (i) the level set of $H_k(\bm p, \bm q)=h_k=\mathrm{const}$ is compact and connected and (ii) the frequencies $\bm\omega=(\omega_1,\dots,\omega_n)$ of quasiperiodic motion with $H(\bm p, \bm q)$  are incommensurate (see below). The first one is a standard assumption in the Liouville-Arnold theorem to show that the dynamics is confined to invariant tori. It
means that the motion is bounded and, roughly speaking, integrals are properly chosen. Consider e.g. a 1D harmonic oscillator $2H =p^2+\omega^2 q^2$.   The level set $H(p, q)=\mathrm{const}$ (ellipse) is connected, so $H$ is a proper choice and \eref{thm} holds.  If we instead  take $\widetilde{H} =(H -h_1)(H -h_2)$ with $h_{1,2}>0$ as our integral,  the manifold $\widetilde{H}(p,q)=\mathrm{const}$ is not always connected.  For example, $ \widetilde{H}=0$ corresponds to two     oscillation amplitudes $A_{1,2}$. The time average of e.g. $q^2$ is either $A_1^2/2$ or $A_2^2/2$ depending on the initial conditions, while the phase space average with $\rho=V^{-1}\delta(\widetilde{H})$ is always  $(A_1^2+A_2^2)/4$.
 
 Let us  go from $(\bm p,\bm q)$ to action-angle variables $(\bm I, \bm \varphi)$. The only information the reader needs   to be able to follow the proof  of \eref{thm} is that this is a canonical transformation, action variables depend only on $H_k$ and vice versa, and   the manifold $H_k(\bm p, \bm q)=h_k$ corresponds to a unique set of values $\alpha_m$ of action variables $I_m$. 
 Hamiltonian equations of motion in action-angle variables are $\dot \varphi_k=\{ H(\bm I), \varphi_k\}=\partial H(\bm I)/\partial I_k\equiv \omega_k$ and $\dot  I_k=\{ H(\bm I),  I_k\}=0$. It follows that $\bm \varphi(t)=\bm \varphi_0+t\bm \omega$. The motion is characterized by $n$ frequencies $\omega_k$ (or $n$ periods).  Incommensurability means  there is no vector $\bm m\ne0$ with integer components such that $\bm m\cdot\bm\omega=0$. The frequencies are incommensurate for most initial conditions in a nondegenerate system, $\det[\partial^2H/\partial I^2]\ne0$\cite{arnold}.
 
 \eref{gme} in new variables reads
\beg
\rho (\bm I)= (2\pi)^{-n}\prod_{k=1}^n \delta\left(I_k-\alpha_{k}\right),
\label{gmeaa}
\en
where we took into account that $\varphi_k$ varies from 0 to $2\pi$ in determining the normalization constant.  Since the Jacobian of a canonical transformation is 1, \eref{thm} becomes
\beg
\begin{split}
\left\langle O(t)\right\rangle_t =\int \! O(\bm I,\bm \varphi)\rho(\bm I)d\bm Id\bm \varphi=
\int \!O(\bm\varphi)\frac{d\bm \varphi}{ (2\pi)^{n}},
\end{split}
\label{thm1}
\en
where on the RHS we supressed the dependence of $O$ on constants $\bm \alpha=(\alpha_1,\dots,\alpha_n)$. 
\eref{thm1} is a known theorem in classical mechanics  called ``the theorem on averages''\cite{arnold}.     We prove it somewhat differently. $O(\bm \varphi)$  is periodic in each $\ph_k$ with period $2\pi$. Expand it in multiple Fourier series
\beg
O(\bm\ph)=\sum_{\bm m} a_{\bm m}e^{2\pi i \bm m\cdot \bm\ph},
\label{ophi}
\en
where the summation is over all $n$-dimensional integer vectors $\bm m$. The time-dependence of $O$ along any trajectory is
\beg
O(t)=O(\bm\ph(t))=\sum_{\bm m} a_{\bm m}e^{2\pi i \bm m\cdot \bm\ph_0}e^{2\pi i t \bm m\cdot \bm\omega}.
\label{ot}
\en
The finite time average of $O(t)$ is
\beg
\frac{1}{T}\int_0^T\!\! O(t) dt=a_0+\sum_{\bm m\ne0} a_{\bm m}\frac{e^{2\pi i T\bm m\cdot \bm\omega -1}}{2\pi i T\bm m\cdot \bm\omega } e^{2\pi i \bm m\cdot \bm\ph_0}.
\label{finite}
\en
 Since $\bm m\cdot \bm\omega\ne 0$ for   $\bm m\ne 0$,    the last equation implies
\beg
\left\langle O(t)\right\rangle_t =\lim_{T\to\infty} \frac{1}{T}\int_0^T\!\! O(t) dt=a_0.
\en
Next we  evaluate the RHS of \eref{thm1} 
\beg
\begin{split}
\int \! O( \bm\varphi)\frac{d\bm \varphi}{ (2\pi)^{n}}=   \sum_{\bm m} a_{\bm m}\prod_{k=1}^n\int_0^{2\pi}  \!   \frac{d\ph_k}{2\pi} e^{2\pi i m_k \ph_k}  \\
= \sum_{\bm m} a_{\bm m} \delta_{m_1 0}\dots  \delta_{m_n 0}=a_0.\\
 \end{split}
 \en
 This completes the proof of theorem \re{gme}. 
 
 Note that the phase space average is unconditionally equal to $a_0$, while the time average equals $a_0$ only when the frequencies are incommensurate. Take, for example,  a 2D anisotropic oscillator $2H =p_1^2+p_2^2 +\omega_1 q_1^2+\omega_2 q_2^2$ and $O =(q_1+q_2)^2$. The phase space and infinite time time averages  are
$(A_1^2+A_2^2)/2$ and $(A_1^2+A_2^2+A_1A_2\cos\alpha\delta_{\omega_1\omega_2})/2$, respectively, where $\alpha$ is the phase shift between the oscillators.  The  averages do not agree when $\omega_1=\omega_2$  and, moreover, the time average depends on an initial condition other than the integrals (on a particular trajectory on the torus).  

As an application of the theorem, consider the semiclassical Dicke model that describes coherent spontaneous emission from a large number of atoms (superradiance) interacting with a cavity electromagnetic mode\cite{dicke,BP}
\beg
H=2\eps S_z +\omega \bar b b+g(\bar b S_-+bS_+),
\label{dicke}
\en
where $\sqrt{2\omega}b=(p-i\omega q)$ is the classical counterpart of the harmonic oscillator annihilation operator and $\vec S$ is a classical spin of length $S$. 
We are interested in the time averaged amplitude of the bosonic mode $\langle |b(t)|\rangle_t$. There are two degrees of freedom and two integrals, $H_1=H$ and $H_2=\bar b b+S_z$, so $H$ is integrable.
In general $|b(t)|$ is an elliptic function. A brute force evaluation of the RHS of \eref{thm} yields a ratio of two complete elliptic integrals. The system has a stable equilibrium at $S_z=-S, b=0$ and an unstable one at $S_z=S, b=0$. Especially interesting are initial conditions in the vicinity of the unstable point. In this case  $|b(t)|$ first grows exponentially and then turns into a sequence of secant pulses (solitons). Let $S_z=S$ and $|b(0)|=r_i$. \eref{thm} now gives
\beg
 \left\langle |b(t)|\right\rangle_t =\frac{\pi r_m}{2\ln(\pi r_m/2r_i)},\quad r_m^2=2S-\frac{(\omega-2\eps)^2}{4g^2},
\en
 where  $r_m$ is the maximum of $|b(t)|$\cite{note3}. 
 
 Finally, we turn to the analysis of $n\to\infty$ limit in \eref{thm}. Since the set of integer vectors $\bm m$ is countable, we can rewrite \eref{ot} as a single sum
 \beg
 O^{(n)}(t)=a_0^{(n)}+\sum_k c_k^{(n)}  e^{i\Omega_k t},
 \label{dis}
 \en
 where we separated the zero frequency term and $a_0^{(n)}=a_0$ in \eref{ophi} assuming the frequencies $\omega_k$ are incommensurate.   The superscript $n$ indicates the number of degrees of freedom. Time averaging first and  then sending $n$ to infinity yields  
\beg
\lim_{n\to\infty} \lim_{T\to\infty} \frac{1}{T}\int_0^T\!\! O^{(n)}(t) dt=\lim_{n\to\infty} a_0^{(n)}\equiv a_0^\infty.
\en
Now consider the opposite order of limits.  In $n\to\infty$ limit the frequency spectrum generally consists of continuum and discrete parts. Nonzero discrete  frequencies do not contribute to the  time average, so we drop them for brevity, 
\beg
O^\infty(t)\equiv\lim_{n\to\infty}  O^{(n)}(t)=a_0^\infty+\int_a^b  e^{i\Omega t} c(\Omega)\nu(\Omega)d\Omega.
\label{ocont}
\en
Whether the limits $T\to\infty$ and $n\to\infty$ commute depends on the behavior of the  function $F(\Omega)=c(\Omega)\nu(\Omega)$ near $\Omega=0$. If $F(\Omega)$ is integrable, the integral in \eref{ocont} vanishes (dephases) as $t\to\infty$ by the Riemann-Lebesgue lemma. Then
\beg
\lim_{n\to\infty}\bigl\langle O^{(n)}(t)\bigr\rangle_t=\bigl\langle O^{\infty}(t)\bigr\rangle_t= a_0^\infty,
\label{comlim}
\en
 i.e. the two limits commute. This is the most likely  scenario in classical integrable 
many-body models. In fact, it is not obvious if there are any reasonable counterexamples.

The relevant  quantity at large $n$ is $O^{\infty}(t)$ and if the limits do not commute, the theorem \re{thm}  looses its predictive power. This happens if $F(\Omega)\propto\delta(\Omega)$ at small $\Omega$.  $\Omega_k$ that vanish in $n\to\infty$ limit must have an anomalously large weight or density $\nu(\Omega)$, so that a  dramatic redistribution from finite to zero $\Omega$ occurs as $n\to\infty$ and a delta-function emerges \textit{from the continuum}. Then the integral in \eref{comlim} does not vanish at $t\to\infty$ and $ O^\infty(t)\not\to a_0^\infty$, while the $n\to\infty$ limit of the time average $\langle O^{(n)}(t)\rangle_t$ is still $a_0^\infty$. 
In linear analysis around the ground state,   $\Omega_k$ are the excitation energies (normal modes). $F(\Omega)\propto\delta(\Omega)$  means  we cannot write their contribution to $O(t)$   in the usual way as an ordinary integral with the density of states. This signals a macroscopic degeneracy of the ground state in the thermodynamic limit.  Similar analysis applies to the order of $N\to\infty$ and $T\to\infty$ limits in \eref{diag}.
Note also that in the absence of isolated nonzero  frequencies  (assuming $F(\Omega)$ is well-behaved) $O^{\infty}(t)\to \mathrm{const}$ at large time. Then the  strong version\cite{note0} of GME holds.  

For example, consider quenches of the detuning $\omega$ in the many-body generalization of the semiclassical Dicke model 
\beg
H =\sum_{k=1}^n 2 \epsilon_k {s}_k^z +\omega{\bar b} {b}  +g\sum_{k=1}^n \left({\bar b} {s}_k^- +b{s}_k^+\right).
\label{Model1}
\en
This is also a classical integrable Hamiltonian\cite{yuzbashyan2014}. It describes an $s$-wave BCS-BEC condensate of atoms and molecules in the mean-field approximation. A relevant observable is the superfluid order parameter $\Delta(t)=gb(t)$. For infinitesimal quenches we know $\Delta(t)$ as well as individual spins $\vec s_i(t)$  at all times and finite $n$. It is straightforward to show that $n\to\infty$ and $T\to\infty$ limits commute. In particular,
\beg
|\Delta(t)|=\Delta_0+\sum_{k=1}^n c_k\cos 2\Omega_k t, 
\label{del}
\en
where $\Omega_k=\sqrt{(\eps_k-\mu)^2+\Delta_0^2}$   up to corrections of order $n^{-1}$ and we suppressed  the superscript $n$. Coefficients $c_k\propto n^{-1}$. In $n\to\infty$ limit the second term in  \eref{del} becomes an integral with a smooth $F(\eps)=c(\eps)\nu(\eps)$ and $|\Delta(t)|\to \Delta_0^\infty=\lim_{n\to\infty}\Delta_0$ as $t\to\infty$, see \cite{yuzbashyan2014} for details. Moreover, the strong version of GME applies.  However, for stronger quenches there is a regime where $|\Delta(t)|$ asymptotes to a periodic function, i.e. discrete nonzero frequencies are present in the continuum limit. Then the strong version no longer holds. The theorem \re{thm} still works and we expect the limits $n\to\infty$ and $T\to\infty$  to commute as before, though the analysis of this case is more difficult.

In conclusion, we pursued two independent threads in this paper. For classical systems, we have seen that the GME given by \eref{thm}   holds for integrable  Hamiltonian systems for a proper choice of integrals as long as the motion is bounded and the frequencies of the quasiperiodic motion are incommensurate. We  argued that the limits of infinite averaging time and number of degrees of freedom typically commute. For quantum systems, we have seen that the GGE hods at any $N$ for maximal $N\times N$ Hamiltonians  that emerge from a rigorous notion of quantum integrability and determined $\beta_k$ in \eref{gge}. The characteristic feature of these systems is that they have the maximum possible number of independent integrals making them   analogous to classical integrable systems. In contrast to  random matrix example we studied, the GGE density matrix for quantum quenches in maximal Hamiltonians   is non-thermal. We discussed the potential reasons of failure of GGE and argued  that one the main reasons is an incomplete set  of integrals as a consequence  of  an ambiguous notion of quantum integrability.

This work was financially supported in part by the David and Lucile Packard Foundation.


\begin{thebibliography}{99}
\bibitem{note0} A stronger version of this statement is that additionally $O(t)$ (expectation value   in the quantum case) goes to a constant as $t\to\infty$. Then the asymptotic value coincides with the time-average and equals the ensemble average.
\end{thebibliography}
\end{document}